\documentclass[12pt,a4paper]{article}

\usepackage{latexsym,feynmf,amsmath}
\usepackage{amsfonts}

\allowdisplaybreaks[2]
\renewcommand{\title}[1]{\null\vspace{6mm}\noindent
                         {\Large{\bf #1}}\vspace{8mm}}
\newcommand{\authors}[1]{\noindent{\large #1}\vspace{0mm}}
\newcommand{\address}[1]{{\center{\noindent\small\itshape #1\vspace{-1mm}}}}

\makeatletter
\def\section{\@startsection{section}{1}{\z@}{-3.25ex plus -1ex minus
             -.2ex}{1.5ex plus .2ex}{\normalfont\bfseries}}
\def\subsection{\@startsection{subsection}{1}{\z@}{-3.25ex plus -1ex
                minus -.2ex}{1.5ex plus .2ex}{\normalfont\itshape}}

\renewenvironment{thebibliography}[1]
         {\section*{References}\frenchspacing\small
          \begin{list}{[\arabic{enumi}]}
         {\usecounter{enumi}\parsep=2pt\topsep 0pt
         \settowidth{\labelwidth}{[#1]}
         \leftmargin=\labelwidth\advance\leftmargin\labelsep
         \rightmargin=0pt\itemsep=0pt\sloppy}}{\end{list}}

\makeatother

\begin{document}

\begin{fmffile}{ncsym_graph}
\begin{titlepage}

\begin{center}
\hspace*{\fill}{{\normalsize \begin{tabular}{l}
                              {\sf hep-th/0203141}\\
                              {\sf TUW 02-05}\\
                              {\sf UWThPh-2002-09}\\
                              {\sf CERN-TH/2002-051}
\end{tabular}   }}

\title{Non-commutative $U(1)$ Super-Yang--Mills Theory: 
\\[0.5ex]
Perturbative Self-Energy Corrections}

\authors{\large{A.~A.~Bichl$^{1\#}$, M. Ertl$^{2*}$, A. Gerhold$^{3*}$, 
J.~M.~Grimstrup$^{4*}$,
H.~Grosse$^{5\&}$,\\[1mm]
L.~Popp$^{6*}$, V. Putz$^{7\$}$, M.~Schweda$^{8*}$, R.~Wulkenhaar$^{9\$}$ }}

\address{$^{\#}$ Theory Division, CERN \\
CH-1211 Geneva 23, Switzerland}
\address{$^*$  Institut f\"ur Theoretische Physik,
Technische Universit\"at Wien \\
Wiedner Hauptstra\ss e 8--10, A-1040 Vienna, Austria}
\address{$^{\&}$  Institut f\"ur Theoretische Physik,
Universit\"at Wien\\
Boltzmanngasse 5, A-1090 Vienna, Austria}
\address{$^{\$}$  Max-Planck-Institute for Mathematics in the Sciences\\
Inselstra\ss{}e 22--26, D-04103 Leipzig, Germany}

\footnotetext[1]{Andreas.Bichl@cern.ch, work supported in
part by ``Fonds zur F\"orderung der Wissenschaftlichen Forschung''
(FWF) under contract P14639-TPH.}

\footnotetext[2]{ertl@tph.tuwien.ac.at, work supported by 
``Fonds zur F\"orderung der Wissenschaftlichen Forschung''
(FWF) under contract P13125-TPH.}

\footnotetext[3]{gerhold@hep.itp.tuwien.ac.at, work supported in
part by ``Fonds zur F\"orderung der Wissenschaft\-lichen Forschung''
(FWF) under contract P13126-TPH.}

\footnotetext[4]{jesper@hep.itp.tuwien.ac.at, work supported by
the Danish Research Agency.}

\footnotetext[5]{grosse@doppler.thp.univie.ac.at.}

\footnotetext[6]{popp@hep.itp.tuwien.ac.at, work supported in part
by ``Fonds zur F\"orderung der Wissenschaftlichen Forschung'' (FWF)
under contract P13125-PHY.}

\footnotetext[7]{volkmar.putz@mis.mpg.de, work supported in
part by ``Fonds zur F\"orderung der Wissenschaftlichen Forschung''
(FWF) under contract P13126-TPH.}

\footnotetext[8]{mschweda@tph.tuwien.ac.at.}

\footnotetext[9]{raimar.wulkenhaar@mis.mpg.de, Schloe\ss{}mann Fellow.}

\vspace{10mm}

\begin{minipage}{12cm}
  {\it Abstract.}  The quantization of the non-commutative $\mathcal{N}
  =1$, $U(1)$ super-Yang--Mills action is performed in the superfield
  formalism. We calculate the
  one-loop corrections to the self-energy of the vector superfield.
  Although the power-counting theorem predicts quadratic ultraviolet
  and infrared divergences, there are actually only logarithmic UV and
  IR divergences, which is a crucial feature of non-commutative
  supersymmetric field theories.
  \vspace*{1cm}
\end{minipage}

\end{center}
\end{titlepage}

$~$\hfill \textit{Dedicated to Olivier Piguet on the occasion of his
  60\/$^{th}$ birthday}

\section {Introduction}
We know that the concept of space-time as a differentiable manifold
cannot be reasonably applied to extremely short distances
\cite{Snyder:1946qz}. Simple heuristic arguments show that it is
impossible to locate a particle with arbitrarily small uncertainty
\cite{Doplicher:tu}.  An interesting concept in order to replace
standard differential geometry is \emph{non-commutative geometry}
pioneered by Connes \cite{Connesbook, Connes:2000by}.
Non-commutative geometry can be regarded as an extension of the
principles of quantum mechanics to geometry itself: space-time
coordinates become non-commutative operators.

The general strategy in non-commutative geometry is to generalize the
mathematical structures encountered in ordinary physics. Standard
quantum field theories deal with problems of interactions at short
distances.  Quantum field theory (QFT) on spaces with different
short-distance structure may therefore show interesting features.
Since singularities in standard QFT are a consequence
of point-like interactions, there has been hope that `smearing out
the points' \cite{Madore} avoids these UV divergences.  However, it
was first noticed by Filk \cite{Filk:dm} that divergences are not
avoided on non-commutative $\mathbb{R}^4$.  This raised the question of
whether the QFT is renormalizable, or not.

Scalar field theories were investigated in \cite{Minwalla:1999px,VanRaamsdonk:2000rr},
where a crucial feature of non-commutative field theories appears---the
{\em UV/IR mixing}.
On the one-loop level the question of renormalizability was
affirmed for Yang--Mills theory on non-commutative $\mathbb{R}^4$ 
\cite{Martin:1999aq,Grosse:2000yy,Armoni:2000xr} and the non-commutative 4-torus 
\cite{Krajewski:1999ja} as well as for
supersymmetric Yang--Mills theory in $(2{+}1)$ dimensions, with space
being the non-commutative 2-torus \cite{Sheikh-Jabbari:1999iw}. QED on
non-commutative $\mathbb{R}^4$ was treated in \cite{Hayakawa:1999yt,Hayakawa:1999zf} and 
BF--Yang--Mills theory in \cite{Benaoum:1999ca}. The Chern--Simons model on
non-commutative space was treated in \cite{Bichl:2000bq}, see also
\cite{Das:2001kf}.

Concerning supersymmetry, also a deformation of the anticommutator of
the fermionic superspace coordinates was considered
\cite{Ferrara:2000mm}, but this deformation is not compatible with
supertranslations and chiral fields.
At the component level, renormalizability of the Wess--Zumino model
to all orders of perturbation theory was shown in \cite{Girotti:2000gc}. 
A superspace formulation (at the
classical level) of the Wess--Zumino model and of super-Yang--Mills
theory was given in \cite{Terashima:2000xq}.
Eventually, renormalizability of the Wess--Zumino model in the
non-commutative superspace formalism was established in \cite{Bichl:2000zu}.
Non-commutative
$\mathcal{N}=1,2$ super-Yang--Mills theories were studied by Zanon in
\cite{Zanon:2000nq}, using the background field method, with the result
that at one loop there are only logarithmic divergences in the
self-energy. This is remarkable because the power-counting theorem
predicts \emph{quadratic divergences} for $\mathcal{N}=1$
super-Yang--Mills theory, which would lead, according to the power-counting analysis of
non-commutative field theories by Chepelev and Roiban
\cite{Chepelev:2000hm}, to non-renormalizability on
non-commutative space-time. The lowering of the degree of divergence
from quadratic to logarithmic seems to be governed by
non-renormalization theorems, see \cite{Kraus:2001tg}.

In this paper we reinvestigate the question of UV/IR mixing in
non-commutative $\mathcal{N}=1$ super-Yang--Mills theory, where we work in the
non-commutative superfield formalism \cite{Bichl:2000zu}.
It turns out that the one-loop self-energy of the
superfield suffer indeed only from \emph{logarithmic} IR divergences.
UV divergences are multiplicatively renormalizable as usual.
Assuming that this behaviour continues to all orders,
non-commutative $\mathcal{N}=1$ super-Yang--Mills theory would be renormalizable, according
to \cite{Chepelev:2000hm}, provided that commutants-type divergences are absent.

Therefore, non-commutativity does not spoil the cancellation of
quadratic and linear divergences in supersymmetric theories, as stated
already in the literature \cite{Matusis:2000jf,Girotti:2000gc,
Khoze:2000sy,Zanon:2000nq,Ruiz:2000hu,VanRaamsdonk:2001jd,Armoni:2001uw}.

On the other hand, non-commutative non-supersymmetric theories
suffer from quadratic (linear) IR divergences which would prevent
renormalizability at higher loop order. Possible ways out
could be hard non-commutative loops resummation \cite{Griguolo:2001wg}
or the use of field redefinitions \cite{Grimstrup:2002nr}.

The paper is organized as follows: Section~\ref{sec2} presents the
Moyal product applied to superfields, while section~\ref{sec3} treats the
action of our model. In section~\ref{sec4} the Legendre transformation
and the perturbative expansion are performed and, after a short
power-counting argument given in section~\ref{sec5}, the self-energy of the
vector superfield is calculated at the one-loop level
(section~\ref{sec6}). Appendices
contain some calculations and conventions.
\section {Moyal Product for Superfields}
\label{sec2}
We consider a non-commutative ($\mathcal{N}=1$) superspace characterized
by the algebra
\begin{equation}
 [\hat x^\mu,\hat x^\nu]=i\Theta^{\mu\nu} \label{a1},
\end{equation}
where $\Theta^{\mu\nu}$ is an antisymmetric, constant and real matrix.
We do not deform the anticommuting coordinates $\theta_\alpha$ and
$\bar\theta^{\dot\alpha}$, i.e.~we assume
\begin{equation}
 \{\theta_\alpha,\theta_\beta\}=\{\bar\theta^{\dot\alpha},\bar\theta^{\dot\beta}\}=
 \{\theta_\alpha,\bar\theta^{\dot\alpha}\}=[\hat x^\mu,\theta_\alpha]=
 [\hat x^\mu,\bar\theta^{\dot\alpha}]=0. \label{a402}
\end{equation}
The non-commutative algebra is represented on an ordinary manifold by
the Moyal product \cite{Filk:dm}. The Moyal product of two vector
superfields can be written as \cite{Bichl:2000zu}
\begin{align}
(\phi\star\phi^\prime)(x,\theta_1,\bar\theta_1)
&= \int dP_{V2}\,dP_{V3}\,\tilde\delta_V(1,2)
\tilde\delta_V(1,3)
\nonumber
\\
&\times\,\tilde\phi(p_2,\theta_2,\bar\theta_2)
 \tilde\phi^\prime(p_3,\theta_3,\bar\theta_3)e^{-i(p_2+p_3)x}
e^{-ip_2\wedge p_3}. \label{a405}
\end{align}
The Moyal product has the important property
\begin{equation}
 \int dV_1\,(\phi\star\phi^\prime)(1)= \int dV_1\,(\phi^\prime\star\phi)(1)=
 \int dV_1\,\phi(1)\phi^\prime(1). \label{a406}
\end{equation}
This implies in particular that one can perform cyclic rotations of
the fields under the integral.

For definiteness we have used vector superfields in (\ref{a405}) and
(\ref{a406}). Of course, one can easily write down analogous formulae
for (anti-)chiral superfields.
\section {The Action}
\label{sec3}
For simplicity we choose the gauge group $U(1)$.  We introduce a
vector superfield $\phi$ whose gauge transformation is given by
\cite{Terashima:2000xq}:
\begin{equation}
 (e^{\phi^\prime})_\star=(e^{-i\bar{\Lambda}})_\star\star(e^\phi)_\star\star(e^{i\Lambda})_\star ,\label{a408}
\end{equation}
with a chiral superfield $\Lambda$ (gauge parameter). 
With the help of the Baker--Campbell--Hausdorff formula we obtain the 
infinitesimal gauge transformation of $\phi$ itself:
\begin{equation}
 \phi^\prime=\phi+i(\Lambda-\bar{\Lambda})+{i\over2}[\phi,\Lambda+\bar{\Lambda}]_\star
 +{i\over12}[\phi,[\phi,\Lambda-\bar{\Lambda}]_\star]_\star+\ldots,
 \label{a413}
\end{equation}
where the dots denote terms that contain three or more powers of $\phi$.
The gauge-invariant NCSYM action is given by
\begin{equation}
 S_{inv}=-{1\over128g^2}\int dS\,W^\alpha W_\alpha \label{a419},
\end{equation}
with
\begin{equation}
 W_\alpha:=\bar D^2\left((e^{-\phi})_\star\star D_\alpha(e^\phi)_\star\right).
\end{equation}
We perform a Taylor expansion of the integrand, 
\begin{align}
S_{inv}=-{1\over128g^2}\int dV\bigg[ & 
-\phi D^\alpha\bar D^2D_\alpha\phi-
 (\bar D^2D^\alpha\phi)[\phi,D_\alpha\phi]_\star 
\nonumber
\\
&-{1\over3}[\phi,\bar D^2D^\alpha\phi]_\star[\phi,D_\alpha\phi]_\star
 +{1\over4}[\phi,D^\alpha\phi]_\star\bar D^2[\phi,D_\alpha\phi]_\star
 +\mathcal{O}(\phi^5)\bigg]. \label{a431}
\end{align}
In order to prepare the quantization, we introduce a chiral 
superfield $B$ (multiplier field) and two
chiral anticommuting superfields $c_+$ (ghost) and $c_-$ (antighost). 
The BRS transformations are given by: 
\begin{align}
 s\phi &= c_+ - \bar c_+ + {1\over2}[\phi,c_++\bar c_+]_\star
 +{1\over12}[\phi,[\phi,c_+-\bar c_+]_\star]_\star+\ldots\nonumber
\hspace*{-10em}
\\
&=: Q_s(\phi,c_+),\nonumber
\\
sc_+ &= -c_+\star c_+, & 
s\bar c_+ &= -\bar c_+\star\bar c_+,\nonumber
\\
sc_- &= B, &
s\bar c_- &=\bar B, \nonumber
\\
 sB&= 0, &
 s\bar B&= 0.
\label{a433}
\end{align}
Now we can write down the BRS-invariant total action:
\begin{equation}
S_{tot}=S_{inv}+S_{gf}+S_{\phi\pi},
\end{equation}
where $S_{inv}$ is given by (\ref{a431}) and the gauge fixing and the 
Faddeev--Popov terms are given by \cite{Piguet:ug}:
\begin{align}
 S_{gf}&= -{1\over128}\int dV(B+\bar B)\phi, \label{a442}\\
 S_{\phi\pi}&= {1\over128}\int dV (c_-+\bar c_-)Q_s(\phi,c_+).
\end{align}
Using (\ref{a433}), the Faddeev--Popov term can be rewritten as
\begin{align}
 S_{\phi\pi} = {1\over128}\int & dV \Bigg(
\bar c_-c_+-c_-\bar c_+\nonumber
\\[-1ex]
& + 
 \Big(c_-+\bar c_-\Big)\bigg({1\over2}[\phi,c_++\bar c_+]_\star
 +{1\over12}[\phi,[\phi,c_+-\bar c_+]_\star]_\star+\ldots\bigg)\Bigg).
 \label{a443}
\end{align}
Again the dots denote terms with three or more powers of $\phi$.

In the following we will also include a mass term in the total action,  
\begin{equation}
 S_{mass}={1\over16g^2}\int dVM^2\phi^2,
\end{equation}
in order to avoid an IR divergence in the propagator of the vector superfield.
\section {Generating Functionals}
\label{sec4}
The generating functional of connected Green's functions for the free
theory can be obtained from the bilinear part $S_{bil}$ of $S_{tot}+S_{mass}$
via a Legendre transformation:
\begin{align}
 Z^c_{bil} &= S_{bil} +\int dVJ\phi 
+\int dS\left(J_BB+\eta_-c_++\eta_+c_-\right)
\nonumber 
\\
&+\int d\bar S\left(J_{\bar B}\bar B+\bar\eta_-\bar c_++\bar\eta_+\bar
  c_- \right)\nonumber
\\
&= S_{bil}+\int dP_V\tilde J_{-p}\tilde\phi_p+\int dP_S\left(\tilde J_{B,-p}\tilde B_p
 +\tilde\eta_{-,-p}\tilde c_{+,p}+\tilde\eta_{+,-p}\tilde
 c_{-,p}\right)
\nonumber
\\
& +\int dP_{\bar S}\left(\tilde J_{\bar B,-p}\tilde{\bar B}_p 
+\tilde{\bar\eta}_{-,-p}
 \tilde{\bar c}_{+,p}+\tilde{\bar\eta}_{+,-p}\tilde{\bar c}_{-,p}\right),
\label{source}
\end{align}
where $\phi$, $B$, $\bar B$, $c_\pm$ and $\bar c_\pm$ are replaced by the
inverse solutions of
\begin{align}
{\delta_V S_{bil}\over\delta_V\tilde\phi_{-p}}&= -\tilde J_p,
\nonumber
\\
{\delta_S S_{bil}\over\delta_S \tilde B_{-p}}&= -\tilde J_{B,p}, 
&
{\delta_{\bar S} S_{bil}\over\delta_{\bar S}\tilde{\bar B}_{-p}}
& =-\tilde J_{\bar B,p}, 
\nonumber
\\
{\delta_S S_{bil}\over\delta_S \tilde c_{-,-p}} &= \tilde \eta_{+,p},
&
{\delta_S S_{bil}\over\delta_S \tilde c_{+,-p}} &=\tilde \eta_{-,p},
\nonumber
\\
{\delta_{\bar S} S_{bil}\over\delta_{\bar S} \tilde {\bar c}_{-,-p}}
&=\tilde{\bar\eta}_{+,p},
&
 {\delta_{\bar S} S_{bil}\over\delta_{\bar S} \tilde {\bar c}_{+,-p}}
&=\tilde{\bar\eta}_{-,p}\label{a468}.
\end{align}
This leads to 
\begin{align}
Z^c_{bil} &= \int dP_{V1}\,dP_{V2}\,{1\over2}\tilde J_{-p_1}
 \Delta_{\phi\phi}(1,2)\tilde J_{-p_2}
 +\int dP_{V1}\,dP_{S2}\,\tilde J_{-p_1}\Delta_{\phi B}(1,2)
\tilde J_{B,-p_2} \nonumber
\\
& +\int dP_{V1}\,dP_{\bar S2}\,\tilde J_{-p_1}\Delta_{\phi\bar B}(1,2)
 \tilde J_{\bar B,-p_2}
 +\int dP_{S1}\,dP_{\bar S2}\,\tilde J_{B,-p_1}\Delta_{B\bar B}(1,2)
 \tilde J_{\bar B,-p_2} \nonumber
\\
&-\int dP_{\bar S1}\,dP_{S2}\,\tilde{\bar\eta}_{-,-p_1} \Delta_{\bar c_+c_-}(1,2)
 \tilde\eta_{+,-p_2}
 -\int dP_{\bar S1}\,dP_{S2}\,\tilde{\bar\eta}_{+,-p_1}
\Delta_{\bar c_-c_+}(1,2) \tilde\eta_{-,-p_2}, 
\end{align}
where the propagators are given by
\begin{align}
\Delta_{\phi\phi}(1,2) &= -g^2(2\pi)^4\delta^4(p_1{+}p_2)e^{p_{2,\mu}
 (\theta_2\sigma^\mu\bar\theta_1-\theta_1\sigma^\mu\bar\theta_2)}
 {1-{1\over4}\theta_{21}^2\bar\theta_{21}^2p_2^2\over
 p_2^2(p_2^2-M^2)}, 
\nonumber
\\
\Delta_{\phi B}(1,2) &= 8(2\pi)^4\delta^4(p_1{+}p_2)
 e^{p_{2,\mu}(\theta_2\sigma^\mu\bar\theta_1-\theta_1\sigma^\mu\bar\theta_2)}
 {1-(\theta_{21}\sigma^\rho\bar\theta_{21})p_{2,\rho}+{1\over4}\theta_{21}^2\bar\theta_{21}^2
 p_2^2\over p_2^2}, 
\nonumber
\\
\Delta_{\phi\bar B}(1,2) &= 8(2\pi)^4\delta^4(p_1{+}p_2)
 e^{p_{2,\mu}(\theta_2\sigma^\mu\bar\theta_1-\theta_1\sigma^\mu\bar\theta_2)}
 {1+(\theta_{21}\sigma^\rho\bar\theta_{21})p_{2,\rho}+{1\over4}\theta_{21}^2\bar\theta_{21}^2
 p_2^2\over p_2^2},
\nonumber
\\
\Delta_{\bar c_+c_-}(1,2) &= \Delta_{\bar c_-c_+}(1,2) 
= \Delta_{\phi B}(1,2),
\nonumber
\\
\Delta_{B\bar B}(1,2) &= {16M^2\over g^2}\Delta_{\phi\bar B}.
\end{align}

The generating functional of general Green's functions is given by
\begin{equation}
 Z={\mathcal N^{-1}}e^{{i\over\hbar}\Gamma_{int}} e^{{i\over\hbar}Z^c_{bil}},
\end{equation}
where $\mathcal N$ is a normalization factor such that $Z[0]=1$ and
\begin{align}
\Gamma_{int}
&= {1\over 4!}\left({\hbar\over i}\right)^{\!4} \int 
dP_{V1}\,dP_{V2}\,dP_{V3}\,dP_{V4}\,\Gamma_{\phi^4}(1,2,3,4)
 {\delta_V^4\over\delta_V\tilde J_{-p_1}\delta_V\tilde J_{-p_2}
\delta_V\tilde J_{-p_3}
 \delta_V\tilde J_{-p_4}}\nonumber
\\
& + {1\over3!}\left({\hbar\over i}\right)^{\!3} \int 
dP_{V1}\,dP_{V2}\,dP_{V3}\,\Gamma_{\phi^3}(1,2,3){\delta_V^3\over
 \delta_V\tilde J_{-p_1}\delta_V\tilde J_{-p_2}\delta_V\tilde
 J_{-p_3}} \nonumber
\\
&+ \left({\hbar\over i}\right)^{\!3} \int 
dP_{V1}\,dP_{V2}\,dP_{V3}\,\Gamma_{c_-c_+\phi}(1,2,3){\delta_S^2
 \delta_V\over\delta_S\tilde\eta_{-,-p_2}\delta_S\tilde\eta_{+,-p_1}
\delta_V\tilde J_{-p_3}} \nonumber
\\
& + \left({\hbar\over i}\right)^{\!3} \int 
dP_{V1}\,dP_{V2}\,dP_{V3}\,\Gamma_{\bar c_-c_+\phi}(1,2,3)
 {\delta_S\delta_{\bar S}\delta_V\over\delta_S\tilde\eta_{-,-p_2}
\delta_{\bar S}
 \tilde{\bar\eta}_{+,-p_1}\delta_V\tilde J_{-p_3}} \nonumber
\\
& + \left({\hbar\over i}\right)^{\!3}\int 
dP_{V1}\,dP_{V2}\,dP_{V3}\,\Gamma_{c_-\bar c_+\phi}(1,2,3)
 {\delta_{\bar S}\delta_S\delta_V\over 
\delta_{\bar S}\tilde{\bar\eta}_{-,-p_2}
 \delta_S\tilde\eta_{+,-p_1}\delta_V\tilde J_{-p_3}} \nonumber
\\
&+ \left({\hbar\over i}\right)^{\!3}\int 
dP_{V1}\,dP_{V2}\,dP_{V3}\,\Gamma_{\bar c_-\bar c_+\phi}(1,2,3)
 {\delta_{\bar S}^2\delta_V\over\delta_{\bar S}
\tilde{\bar\eta}_{-,-p_2}\delta_{\bar S}
 \tilde{\bar\eta}_{+,-p_1}\delta_V\tilde J_{-p_3}} \nonumber
\\
&+ {1\over2!}\left({\hbar\over i}\right)^{\!4}\int 
dP_{V1}\,dP_{V2}\,dP_{V3}\,dP_{V4}\,\Gamma_{c_-c_+\phi^2}(1,2,3,4)
 {\delta_S^2\delta_V^2\over\delta_S\tilde\eta_{-,-p_2}
\delta_S\tilde\eta_{+,-p_1}\delta_V\tilde J_{-p_3}
 \delta_V\tilde J_{-p_4}} \nonumber
\\
& + {1\over2!}\left({\hbar\over i}\right)^{\!4}\int 
dP_{V1}\,dP_{V2}\,dP_{V3}\,dP_{V4}\,\Gamma_{\bar c_-c_+\phi^2}
 (1,2,3,4){\delta_S\delta_{\bar S}\delta_V^2\over
\delta_S\tilde\eta_{-,-p_2}\delta_{\bar S}
 \tilde{\bar\eta}_{+,-p_1}\delta_V\tilde J_{-p_3}
\delta_V\tilde J_{-p_4}}\nonumber
\\
& + {1\over2!}\left({\hbar\over i}\right)^{\!4}\int 
dP_{V1}\,dP_{V2}\,dP_{V3}\,dP_{V4}\,\Gamma_{c_-\bar c_+\phi^2}
 (1,2,3,4){\delta_{\bar S}\delta_S\delta_V^2\over
\delta_{\bar S}\tilde{\bar\eta}_{-,-p_2}\delta_S
 \tilde\eta_{+,-p_1}\delta_V\tilde J_{-p_3}\delta_V\tilde J_{-p_4}} 
\nonumber
\\
& + {1\over2!}\left({\hbar\over i}\right)^{\!4}\int 
dP_{V1}\,dP_{V2}\,dP_{V3}\,dP_{V4}\,\Gamma_{\bar c_-\bar c_+\phi^2}
 (1,2,3,4){\delta_{\bar S}^2\delta_V^2\over
\delta_{\bar S}\tilde{\bar\eta}_{-,-p_2}\delta_{\bar S}
 \tilde{\bar\eta}_{+,-p_1}\delta_V\tilde J_{-p_3}
\delta_V\tilde J_{-p_4}},
\end{align}
where we have defined
\begin{align}
\Gamma_{\phi^3}(1,2,3) =& 
{\delta_V^3S_{int}\over\delta_V\tilde\phi_{p_1}\delta_V\tilde\phi_{p_2}
 \delta_V\tilde\phi_{p_3}}\bigg|_0, 
\nonumber
\\
 \Gamma_{\phi^4}(1,2,3,4) &= 
{\delta_V^4S_{int}\over\delta_V\tilde\phi_{p_1}\delta_V\tilde\phi_{p_2}
 \delta_V\tilde\phi_{p_3}\delta_V\tilde\phi_{p_4}}\bigg|_{0}, 
\nonumber
\\
 \Gamma_{c_-c_+\phi}(1,2,3)&= {\delta_V^3S_{int}\over\delta_V
\tilde c_{-,p_1}\delta_V\tilde c_{+,p_2}
 \delta_V\tilde\phi_{p_3}}\bigg|_0,
\nonumber
\\
 \Gamma_{c_-c_+\phi^2}(1,2,3,4)&= {\delta_V^4S_{int}
\over\delta_V\tilde c_{-,p_1}
 \delta_V\tilde c_{+,p_2}\delta_V\tilde\phi_{p_3}\delta_V
\tilde\phi_{p_4}}\bigg|_0,
\label{a4156}
\end{align}
and similarly for the terms $\Gamma_{\bar c_-c_+\phi}(1,2,3)$, $\Gamma_{c_-\bar
  c_+\phi}(1,2,3)$, $\Gamma_{\bar c_-\bar c_+\phi}(1,2,3)$,
$\Gamma_{\bar c_-c_+\phi^2}(1,2,3,4)$, $\Gamma_{c_-\bar
  c_+\phi^2}(1,2,3,4)$ and $\Gamma_{\bar c_-\bar c_+\phi^2}(1,2,3,4)$.
Here, $S_{int}$ is the interaction part of $S_{tot}$, and the
subscript $0$ means that all the fields have to be set to zero
after the functional derivatives have been performed. The mixture of
(anti-)chiral and vectorial field derivatives in the ghost sector is
due to our convention that source terms for the ghosts involve the
(anti-)chiral measure (\ref{source}), whereas interactions between
ghosts and vector superfields are defined in terms of the vectorial measure
(\ref{a443}). One should notice here that these are all necessary
vertices for the calculation of the one-loop self-energy part of the
vector superfield. The final results for these vertex functions
(\ref{a4156}) are rather complicated and can be looked up in appendix
A.

Furthermore, we mention the generating functional of connected Green's
functions, which is given by
\begin{equation}
 Z^c = {\hbar\over i}\ln Z. \label{z}
\end{equation}
\section{Power-Counting}
\label{sec5}
We note that, apart from the exponentials and the $\theta$-factors in the 
numerator, the vector field propagators are of order ${1\over (p^2)^2}$ 
and the ghost propagators of order ${1\over p^2}$.
We consider the exponentials and the $\theta$-factors. 
From the invariance of Green's functions with respect to translations 
and supersymmetry transformations one finds that a 
one-particle irreducible Green's function in momentum space can always
be written as \cite{Piguet:ug}:
\begin{align} 
 \tilde \Gamma(1,\ldots,n) &= {\delta^n \Gamma[\phi_i]\over 
{\delta\tilde\phi(p_1)
 \ldots\delta\tilde\phi(p_n)}}\nonumber
\\ 
&= (2\pi)^4\delta^4(\sum_{j=1}^n p_j)e^{-\sum_{i=2}^n p_{i,\mu}
(\theta_i\sigma^\mu
 \bar\theta_1-\theta_1\sigma^\mu\bar\theta_i)}\tilde f(-p_2,...,-p_n,
 \theta_{i1},\bar\theta_{i1}).\quad \label{lll}
\end{align}
This general structure is true in particular for propagators and
vertices. Thus, the general structure of the integrand of the
superspace integral corresponding to an arbitrary Feynman graph is
 \begin{equation}
 I = \exp\Big(\underbrace{-\sum_{i\leq j}\sum_\tau l_{ij,\tau,\mu}
 (\theta_i\sigma^\mu\bar\theta_j-\theta_j\sigma^\mu\bar\theta_i)}_{
 E(p,k,\theta_l,\bar\theta_l)}\Big)
 \prod_{i\leq j}\prod_\tau\tilde f_{ij,\tau}
 (-l_{ij,\tau},\theta_{ij},\bar\theta_{ij}).
 \end{equation}
Here, $l_{ij,\tau}$ is the momentum running from point $i$ to point $j$, 
and $\tau$ counts momenta running between the same pair of points.
We have chosen a basis for $(l_{ij,\tau}) =(p,k)$, where $p$ and $k$ are 
the external and internal momenta, respectively. With momentum 
conservation,
 \begin{equation}
 \sum_{j,\tau}l_{ij,\tau} = p_i,
 \end{equation}
we find 
 \begin{equation}
 E(p,k,\theta_l,\bar\theta_l) = E(p,k,\theta_{l1},\bar\theta_{l1})
 -\sum_i(\theta_i\sigma^\mu\bar\theta_1-\theta_1\sigma^\mu\bar\theta_i)p_{i,\mu}.
 \end{equation}   
Therefore, the exponentials appearing in the formulae for the 
propagators and vertices can be rewritten as
 \begin{equation}
 e^{k_\mu(\theta_i\sigma^\mu
 \bar\theta_j-\theta_j\sigma^\mu\bar\theta_i)} \Rightarrow
 e^{k_\mu(\theta_{i1}\sigma^\mu
 \bar\theta_{j1}-\theta_{j1}\sigma^\mu\bar\theta_{i1})},
 \end{equation}
if and only if $k$ is an internal momentum. A Taylor expansion of the exponentials
shows that from the $\theta$-factors and the exponential we will at most get terms 
like $\theta_{i1}^2\bar\theta_{i1}^2 k^2$. The highest power of $k^2$ that 
can appear is just the number of independent differences $\theta_{ij}$ (with
$j=1$ in the calculation above) that can be constructed, which is exactly $n-1$,
$n$ being the number of vertices. So we find for the superficial divergence 
degree of an 1PI-graph
 \begin{equation}
 d(\Gamma) = 4L-2G-4V+2(n_G+n_V-1)+2n_V. \label{b58}
 \end{equation}
Here, $L$ is the number of loop integrations, $G$ and $V$ are the numbers 
of ghost and vector superfield propagators, respectively, $n_G$ and $n_V$
count the ghost-vector superfield and the pure-vector superfield vertices.
The last term, $2n_V$, has to be included in (\ref{b58}) because of the four covariant derivatives that appear in the parts of the 
Lagrangian corresponding to the three and four vertices of the vector superfield.

Using the topological relation $L= G+V-n_G-n_V+1$ and charge 
conservation for the ghost fields, $2n_G = 2G+N_G$ ($N_G$ being 
the number of external ghost fields), we find
 \begin{equation}
 d(\Gamma) = 2-N_G.
 \end{equation}
For the vector superfield self-energy ($N_G =0$), this means a 
superficial degree of divergence of 2. 
\section {Self-Energy of the Vector Superfield}
\label{sec6}
The following Feynman graphs contribute to the self-energy of the
vector superfield at the one-loop level (continuous lines denote
vector superfield propagators, dotted lines ghost propagators):
\begin{align}
I_1:\quad & \parbox{40mm}{
 \begin{fmfgraph}(40,20)
  \fmfleft{l}
  \fmfright{r}
  \fmf{plain}{l,i1,r}
  \fmf{plain,right=2}{i1,i1}
  \fmfdot{i1}
 \end{fmfgraph}}
 \put(-37,2){\small$1$}
 \put(-28,5){\small$4$}
 \put(-15,5){\small$3$}
 \put(-6,2){\small$2$}
&
I_2:\quad & \parbox{40mm}{
 \begin{fmfgraph}(40,20)
  \fmfleft{l}
  \fmfright{r}
  \fmf{plain}{l,i1,r}
  \fmf{dots,right=2}{i1,i1}
  \fmfdot{i1}
 \end{fmfgraph}}
 \put(-37,2){\small$1$}
 \put(-28,5){\small$4_+$}
 \put(-15,5){\small$\bar{3}_-$}
 \put(-6,2){\small$2$}
\nonumber
\\
I_3:\quad & \parbox{40mm}{
 \begin{fmfgraph}(40,20)
  \fmfleft{l}
  \fmfright{r}
  \fmf{plain}{l,i1,r}
  \fmf{dots,right=2}{i1,i1}
  \fmfdot{i1}
 \end{fmfgraph}}
 \put(-37,2){\small$1$}
 \put(-28,5){\small$\bar 4_+$}
 \put(-15,5){\small$3_-$}
 \put(-6,2){\small$2$}
&
I_4:\quad & \parbox{40mm}{
 \begin{fmfgraph}(40,20)
  \fmfleft{l}
  \fmfright{r}
  \fmf{plain,tension=2.5}{l,i1}
  \fmf{plain,right}{i1,i2}
  \fmf{plain,right}{i2,i1}
  \fmf{plain,tension=2.5}{i2,r}
  \fmfdot{i1,i2}
 \end{fmfgraph}}
 \put(-37,2){\small$1$}
 \put(-30,5){\small$4$}
 \put(-30,-5){\small$3$}
 \put(-12,5){\small$5$}
 \put(-12,-5){\small$6$}
 \put(-6,2){\small$2$}
 \nonumber
\\
I_5:\quad & \parbox{40mm}{
 \begin{fmfgraph}(40,20)
  \fmfleft{l}
  \fmfright{r}
  \fmf{plain,tension=2.5}{l,i1}
  \fmf{dots,right}{i1,i2}
  \fmf{dots,right}{i2,i1}
  \fmf{plain,tension=2.5}{i2,r}
  \fmfdot{i1,i2}
 \end{fmfgraph}}
 \put(-37,2){\small$1$}
 \put(-32,5){\small$4_+$}
 \put(-32,-5){\small$3_-$}
 \put(-12,5){\small$\bar{5}_-$}
 \put(-12,-5){\small$\bar{6}_+$}
 \put(-6,2){\small$2$}
&
I_6:\quad & \parbox{40mm}{
 \begin{fmfgraph}(40,20)
  \fmfleft{l}
  \fmfright{r}
  \fmf{plain,tension=2.5}{l,i1}
  \fmf{dots,right}{i1,i2}
  \fmf{dots,right}{i2,i1}
  \fmf{plain,tension=2.5}{i2,r}
  \fmfdot{i1,i2}
 \end{fmfgraph}}
 \put(-37,2){\small$1$}
 \put(-32,5){\small$\bar{4}_+$}
 \put(-32,-5){\small$\bar{3}_-$}
 \put(-12,5){\small$5_-$}
 \put(-12,-5){\small$6+$}
 \put(-6,2){\small$2$}
 \nonumber
\\
I_7:\quad & \parbox{40mm}{
 \begin{fmfgraph}(40,20)
  \fmfleft{l}
  \fmfright{r}
  \fmf{plain,tension=2.5}{l,i1}
  \fmf{dots,right}{i1,i2}
  \fmf{dots,right}{i2,i1}
  \fmf{plain,tension=2.5}{i2,r}
  \fmfdot{i1,i2}
 \end{fmfgraph}}
 \put(-37,2){\small$1$}
 \put(-32,5){\small$\bar4_+$}
 \put(-32,-5){\small$3_-$}
 \put(-12,5){\small$5_-$}
 \put(-12,-5){\small$\bar{6}_+$}
 \put(-6,2){\small$2$}
&
I_8:\quad & \parbox{40mm}{
 \begin{fmfgraph}(40,20)
  \fmfleft{l}
  \fmfright{r}
  \fmf{plain,tension=2.5}{l,i1}
  \fmf{dots,right}{i1,i2}
  \fmf{dots,right}{i2,i1}
  \fmf{plain,tension=2.5}{i2,r}
  \fmfdot{i1,i2}
 \end{fmfgraph}}
 \put(-37,2){\small$1$}
 \put(-32,5){\small$4_+$}
 \put(-32,-5){\small$\bar 3_-$}
 \put(-12,5){\small$\bar 5_-$}
 \put(-12,-5){\small$6_+$}
 \put(-6,2){\small$2$}
\end{align}
{}From the generating functional (\ref{z}) we obtain the following integrals corresponding to these
graphs (after amputation of the external lines):
\begin{align}
 I_1&= {\hbar\over2i}\int dP_{V3}dP_{V4}\,
\Gamma_{\phi^4}(3,4,1,2)\Delta_{\phi\phi}(3,4),
\nonumber
\\ 
I_2&= -{\hbar\over i}\int dP_{V3}dP_{V4}\,
\Gamma_{\bar c_-c_+\phi^2}(3,4,1,2)\Delta_{\bar c_-c_+}(3,4),
\nonumber
\\
I_3&= -{\hbar\over i}\int dP_{V3}dP_{V4}\,
\Gamma_{c_-\bar c_+\phi^2}(3,4,1,2)\Delta_{c_-\bar c_+}(3,4),
\nonumber
\\
I_4&= {\hbar\over2i}\int dP_{V3}dP_{V4}dP_{V5}dP_{V6}\,
 \Gamma_{\phi^3}(3,4,1)\Delta_{\phi\phi}(3,6)\Gamma_{\phi^3}(5,6,2)
\Delta_{\phi\phi}(5,4),
\nonumber
\\
I_5&= -{\hbar\over i}\int dP_{V3}dP_{V4}dP_{V5}dP_{V6}\,
\Gamma_{c_-c_+\phi}(3,4,1)\Delta_{c_-\bar c_+}(3,6)
\Gamma_{\bar c_-\bar c_+\phi}(5,6,2)\Delta_{\bar c_-c_+}(5,4), 
\nonumber
\\
I_6&= -{\hbar\over i}\int dP_{V3}dP_{V4}dP_{V5}dP_{V6}\,
\Gamma_{\bar c_-\bar c_+\phi}(3,4,1) \Delta_{\bar c_-c_+}(3,6)
\Gamma_{c_-c_+\phi}(5,6,2)\Delta_{c_-\bar c_+}(5,4),
\nonumber
\\
I_7&= -{\hbar\over i}\int dP_{V3}dP_{V4}dP_{V5}dP_{V6}\,
\Gamma_{c_-\bar c_+\phi}(3,4,1)
 \Delta_{c_-\bar c_+}(3,6)
\Gamma_{c_-\bar c_+\phi}(5,6,2)\Delta_{c_-\bar c_+}(5,4),
\nonumber
\\
I_8&= -{\hbar\over i}\int dP_{V3}dP_{V4}dP_{V5}dP_{V6}\,
\Gamma_{\bar c_-c_+\phi}(3,4,1)
 \Delta_{\bar c_-c_+}(3,6) 
\Gamma_{\bar c_-c_+\phi}(5,6,2)\Delta_{\bar c_-c_+}(5,4). \label{p2}
\end{align}
(Note that $\Delta_{c_-\bar c_+}(3,6)= -\Delta_{\bar c_+c_-}(6,3)$.)
We now insert the explicit expressions for the propagators and
vertices into the eight integrals in (\ref{p2}). After some lengthy
simplifications of the integrands (see appendix A) we arrive at
\begin{align}
I_1&= {\hbar\over128i}\delta^4(p_1+p_2) 
e^{-p_{1,\mu}(\theta_1\sigma^\mu\bar\theta_2
 -\theta_2\sigma^\mu\bar\theta_1)}
\int d^4k\,\sin^2(p_1\wedge k){1\over k^2(k^2-M^2)}\nonumber
\\
&\times\left(-{2\over3}
+{1\over6} \theta_{12}^2\bar\theta_{12}^2
 (2k^2+p_1^2)\right),\label{q1}
\\
I_2&= {\hbar\over384i}\delta^4(p_1+p_2)
 e^{-p_{1,\mu}(\theta_1\sigma^\mu\bar\theta_2
 -\theta_2\sigma^\mu\bar\theta_1)}
\int d^4k\,\sin^2(p_1\wedge k){1\over k^2}\theta_{12}^2
 \bar\theta_{12}^2,
\\
I_3&= {\hbar\over384i}\delta^4(p_1+p_2)
 e^{-p_{1,\mu}(\theta_1\sigma^\mu\bar\theta_2
 -\theta_2\sigma^\mu\bar\theta_1)}
\int d^4k\,\sin^2(p_1\wedge k){1\over
  k^2}\theta_{12}^2\bar\theta_{12}^2,
\\
I_4&= {\hbar\over128i}\delta^4(p_1+p_2) 
e^{-p_{1,\mu}(\theta_1\sigma^\mu\bar\theta_2
 -\theta_2\sigma^\mu\bar\theta_1)}
\int d^4k\,\sin^2(p_1\wedge k) \nonumber
\\&\times\,{1\over k^2(k^2-M^2)(k+p_1)^2((k+p_1)^2-M^2)}
 \nonumber
\\
&\times\bigg(-4(k^2)^2-8k^2(kp_1)-5k^2p_1^2-p_1^2(kp_1)
-{1\over4} \theta_{12}^2\bar\theta_{12}^2
 \Big(-4(k^2)^2p_1^2 \nonumber
\\
& \hspace*{5em} + 8k^2(kp_1)^2+ 8(kp_1)^3+8(kp_1)^2p_1^2
+(kp_1)(p_1^2)^2-3(p_1^2)^2k^2\Big)\bigg),
\\
I_5&= {\hbar\over256i}\delta^4(p_1+p_2) 
 e^{-p_{1,\mu}(\theta_1\sigma^\mu\bar\theta_2
 -\theta_2\sigma^\mu\bar\theta_1)}
\left(1-(\theta_{12}\sigma^\rho\bar\theta_{12})p_{1,\rho}+
 {1\over4}\theta_{12}^2\bar\theta_{12}^2p_1^2\right)
\nonumber
\\
&\times\int d^4k\,\sin^2(p_1\wedge k){1\over k^2(p_1+k)^2},
\\
I_6&= {\hbar\over256i}\delta^4(p_1+p_2) 
 e^{-p_{1,\mu}(\theta_1\sigma^\mu\bar\theta_2
 -\theta_2\sigma^\mu\bar\theta_1)}
\left(1+(\theta_{12}\sigma^\rho\bar\theta_{12})p_{1,\rho}+
 {1\over4}\theta_{12}^2\bar\theta_{12}^2p_1^2\right)
\nonumber
\\
&\times\int d^4k\,\sin^2(p_1\wedge k){1\over k^2(p_1+k)^2},
\\
I_7&= -{\hbar\over256i}\delta^4(p_1+p_2) 
 e^{-p_{1,\mu}(\theta_1\sigma^\mu\bar\theta_2
 -\theta_2\sigma^\mu\bar\theta_1)}
\int d^4k\,\sin^2(p_1\wedge k){1\over k^2(p_1+k)^2}
\nonumber
\\
&\times\left(1-(\theta_{12}\sigma^\rho\bar\theta_{12})(p_1+2k)_\rho+
 \theta_{12}^2\bar\theta_{12}^2\left({1\over4}p_1^2+p_1k+k^2
\right)\right), \label{g2}
\\
I_8&= -{\hbar\over256i}\delta^4(p_1+p_2) 
 e^{-p_{1,\mu}(\theta_1\sigma^\mu\bar\theta_2
 -\theta_2\sigma^\mu\bar\theta_1)}
\int d^4k\,\sin^2(p_1\wedge k){1\over k^2(p_1+k)^2}\nonumber
\\
&\times\left(1+(\theta_{12}\sigma^\rho\bar\theta_{12})(p_1+2k)_\rho+
 \theta_{12}^2\bar\theta_{12}^2
\left({1\over4}p_1^2+p_1k+k^2\right)\right).\label{q2}
\end{align} 
This gives up to terms in the integrand, which evaluate for $M \neq 0$
to finite quantities 
\begin{align}
\sum_{i=1}^8 I_i 
&= \frac{\hbar}{128i} \delta^4(p_1+p_2) 
e^{-p_{1,\mu}(\theta_1\sigma^\mu\bar\theta_2
 -\theta_2\sigma^\mu\bar\theta_1)}
\int d^4k\,\frac{\sin^2(p_1\wedge k)}{k^2(k+p_1)^2} 
\nonumber
\\
& 
\times\Big(-{14\over3} - \frac{1}{4} \theta_{12}^2 \bar{\theta}_{12}^2 
\Big( -4 kp_1 - \frac{26}{3} p_1^2
+ \frac{8 (kp_1)^2}{k^2-M^2} \Big) \Big).
\label{final}
\end{align}
As usual we write $\sin^2(p_1 \wedge k)= \frac{1}{2} -
\frac{1}{2}\cos(2 p_1 \wedge k)$ and refer to the part corresponding
to $\frac{1}{2}$ as `planar' and the part corresponding to
$\frac{1}{2}\cos (2 p_1 \wedge k)$ as `non-planar'. The planar part of
(\ref{final}) is UV-divergent and evaluated in dimensional or analytic
regularization to
\begin{align}
\Big(\sum_{i=1}^8 I_i \Big)_{planar}^{reg} &= 
- \frac{7}{3} \frac{\hbar \pi^2 }{128
  \varepsilon} \,  \delta^4(p_1+p_2) 
e^{-p_{1,\mu}(\theta_1\sigma^\mu\bar\theta_2
 -\theta_2\sigma^\mu\bar\theta_1)}
\Big(1 - \frac{1}{4} \theta_{12}^2 \bar{\theta}_{12}^2 p_1^2 \Big)
+ \mathcal{O}(1)\,.
\end{align}
This means that the divergence in the planar part of the self-energy is
transversal, so that it can be removed by multiplicative
renormalization. Because of the oscillating integrand, the non-planar part of (\ref{final}) turns out to be
finite for $p_1 \neq 0$ and is
evaluated to (with $\tilde p_1^\mu := \Theta^{\mu\nu}p_{1,\nu}$): 
\begin{align}
\Big(\sum_{i=1}^8 I_i \Big)_{\text{\textit{non-planar}}}
&= \frac{7}{3} \frac{\hbar \pi^2}{64} \delta^4(p_1+p_2)\,
e^{-p_{1,\mu}(\theta_1\sigma^\mu\bar\theta_2
 -\theta_2\sigma^\mu\bar\theta_1)}
\nonumber
\\
&\times\Big(1-\frac{1}{4}\theta_{12}^2 
\bar{\theta}_{12}^2 p_1^2 \Big) \int_0^1 \!\!\! dx\, 
K_0\Big(\sqrt{x(1{-}x)p_1^2\tilde{p_1}^2}\Big) + \mathcal{O}(1)\,,
\label{IR}
\end{align}
where $\mathcal{O}(1)$ in (\ref{IR}) collects terms that are regular
for $p_1\to 0$, and $K_0(y)=\log 2 -\log y - \gamma + \mathcal{O}(y^2)$ is
the modified Bessel function of second kind (with Euler's $\gamma = 
0.577...$). Thus, the non-planar part
of the self-energy is only logarithmically
divergent for $p_1\to 0$,
as expected for a supersymmetric theory,
despite the fact that it is non-commutative.
\section {Conclusions}
We have computed the one-loop self-energy of the vector superfield in
non-commutative $\mathcal{N}=1$, $U(1)$ super-Yang--Mills theory
in the superfield formulation, where
our results can
be summarized in the following way:
\begin{itemize}
\item UV divergences can be multiplicatively renormalized
in the usual way.
\item IR divergences are only logarithmic and do not spoil
renormalizability.
\end{itemize}
Because there are only
logarithmic divergences, a non-commutative field theory is, according to
\cite{Chepelev:2000hm}, power-counting renormalizable (assuming there
is no problem with commutants). This would imply that non-commutative
$\mathcal{N}=1$, $U(1)$ super-Yang--Mills theory is renormalizable, because
supersymmetry avoids quadratic (and linear) IR divergences.
\section*{Acknowledgements}
A.~A.~B.\ would like to thank Adi Armoni for numerous enlightening discussions.
We would also like to thank V.~V.~Khoze, V.~O.~Rivelles, F.~Ruiz Ruiz and G.~Travaglini
for various useful comments. 
\appendix
\renewcommand{\theequation}{\Alph{section}.\arabic{equation}}
\makeatletter 
\@addtoreset{equation}{section} 
\makeatother
\section {Calculations}
\subsection {$\phi^4$-vertex}
Let us compute the $\phi^4$-vertex given in (\ref{a4156}),
$\displaystyle
\Gamma_{\phi^4}(1,2,3,4)={\delta_V^4S_{\phi^4}\over
\delta_V\tilde\phi_{p_1}\delta_V\tilde\phi_{p_2}
  \delta_V\tilde\phi_{p_3}\delta_V\tilde\phi_{p_4}}$, where
$S_{\phi^4}$ denotes the part of the total action which is quartic in
$\phi$:
\begin{align}
 S_{\phi^4}= -{1\over128g^2}\int dV \bigg({1\over4}[
\bar D^2\phi,D^\alpha\phi]_\star
 & -{1\over12}[\phi,\bar
 D^2D^\alpha\phi]_\star
+{1\over2}[\bar D_{\dot\alpha}\phi,
\bar D^{\dot\alpha}D^\alpha\phi]_\star\bigg)
 [\phi,D_\alpha\phi]_\star. \label{a4162}
\end{align}
Using the definition of the Moyal product (\ref{a405}) and integrating
by parts several times, this expression can be written as
\begin{align}
 S_{\phi^4} =-{1\over256g^2}\int dP_{V5}\,dP_{V6}\,dP_{V7}\,dP_{V8}\,
& (2\pi)^4\delta^4(p_5+p_6+p_7+p_8)
 e^{-ip_5\wedge p_6-ip_7\wedge p_8}
\nonumber
\\
& \times \tilde\phi_{p_5}\tilde\phi_{p_6}\tilde\phi_{p_7}\tilde\phi_{p_8}
 \mathcal V(5,6,7,8) \tilde\delta_V(5,6)\tilde\delta_V(5,7)\tilde\delta_V(5,8), \label{a4165}
\end{align}
where the differential operator $\mathcal V$ is given by
\begin{align}
\mathcal V&(5,6,7,8)\nonumber
\\
&=\left(\tilde D^\alpha_{8,-p_8}-\tilde D^\alpha_{7,-p_7}\right)
\tilde D_{\alpha,6,-p_6}
 \bigg[{1\over2}\tilde{\bar D}^2_{5,-p_5}
-{1\over6}
 \tilde{\bar D}^2_{6,-p_6}+\tilde{\bar D}_{\dot\alpha,6,-p_6}
\tilde{\bar D}^{\dot\alpha}_{5,-p_5}
 \bigg]\nonumber
\\
&+\left(\tilde D^\alpha_{7,-p_7}-\tilde D^\alpha_{8,-p_8}\right)
\tilde D_{\alpha,5,-p_5}
 \bigg[{1\over2}\tilde{\bar D}^2_{6,-p_6}
-{1\over6}
 \tilde{\bar D}^2_{5,-p_5}+\tilde{\bar D}_{\dot\alpha,5,-p_5}
\tilde{\bar D}^{\dot\alpha}_{6,-p_6}
 \bigg].
\end{align}
We notice that $\mathcal V$ has the following symmetry properties:
\begin{equation}
 \mathcal V(5,6,7,8)=-\mathcal V(6,5,7,8)=-\mathcal V(5,6,8,7)=\mathcal V(6,5,8,7).
\end{equation}
Using these symmetries we can write $\Gamma_{\phi^4}$ as
\begin{align}
 \Gamma_{\phi^4}(1,2,3,4)&= {1\over64g^2}(2\pi)^4
\delta^4(p_1+p_2+p_3+p_4) \,
\nonumber
\\
& \times\bigg[\sin(p_1\wedge p_2)\sin(p_3\wedge p_4)\left(\mathcal V(1,2,3,4)
 +\mathcal V(3,4,1,2)\right) \nonumber
\\
&~~ +\sin(p_1\wedge p_3)\sin(p_4\wedge p_2)\left(\mathcal V(1,3,4,2)
 +\mathcal V(4,2,1,3)\right) \nonumber
\\
& ~~ +\sin(p_1\wedge p_4)\sin(p_2\wedge p_3)\left(\mathcal V(1,4,2,3)
 +\mathcal V(2,3,1,4)\right)\bigg] \nonumber 
\\
& \times\tilde\delta_V(1,2)\tilde\delta_V(1,3)\tilde\delta_V(1,4) .
\label{a4171}
\end{align}
In order to simplify this expression we have to evaluate terms like
\begin{equation} 
 A(1,2,3,4):=\tilde D^\alpha_{4,-p_4}\tilde D_{\alpha,2,-p_2}
 \tilde{\bar D}_{\dot\alpha,2,-p_2}\tilde{\bar D}^{\dot\alpha}_{1,-p_1}
 \tilde\delta_V(1,2)\tilde\delta_V(1,3)\tilde\delta_V(1,4).
\label{a4172}
\end{equation}
This can be done most easily if we first insert exponentials in the
following way:
\begin{eqnarray}
 &&\!\!\!\!\!\!\!\!\!\!\!A(1,2,3,4)=\left(\tilde D^\alpha_{4,-p_4}e^{p_{4,\mu}
 (\theta_3\sigma^\mu\bar\theta_4
 -\theta_4\sigma^\mu\bar\theta_3)}\tilde\delta_V(4,3)\right)\nonumber\\
 &&\!\!\!\!\!\!\!\!\!\times\left(\tilde D_{\alpha,2,-p_2} \tilde{\bar D}_{\dot\alpha,2,-p_2}e^{p_{2,\nu}(\theta_3\sigma^\nu\bar\theta_2
 -\theta_2\sigma^\nu\bar\theta_3)}\tilde\delta_V(2,3)\right)
 \left(\tilde{\bar D}^{\dot\alpha}_{1,-p_1} e^{p_{1,\rho}(\theta_3\sigma^\rho\bar\theta_1
 -\theta_1\sigma^\rho\bar\theta_3)}\tilde\delta_V(1,3)\right),\qquad
\end{eqnarray}
which is allowed because of $\theta^3=\bar\theta^3=0$. We now use the identities
\begin{eqnarray}
 &&\tilde D_{\alpha,i,p}e^{p_\mu(\theta_i\sigma^\mu\bar\theta_j-\theta_j\sigma^\mu\bar\theta_i)}=
 e^{p_\mu(\theta_i\sigma^\mu\bar\theta_j-\theta_j\sigma^\mu\bar\theta_i)}
 \tilde D_{\alpha,ij,p}, \label{a4118}\\
 &&\tilde {\bar D}_{\dot\alpha,i,p}e^{p_\mu(\theta_i\sigma^\mu\bar\theta_j
 -\theta_j\sigma^\mu\bar\theta_i)}=
 e^{p_\mu(\theta_i\sigma^\mu\bar\theta_j-\theta_j\sigma^\mu\bar\theta_i)}
 \tilde{\bar D}_{\dot\alpha,ij,p}, \label{a4119}
\end{eqnarray}
which can easily be verified, leading to
\begin{equation}
 A(1,2,3,4)=e^{E_3^{(4)}} \left(\tilde D^\alpha_{43,-p_4}\tilde\delta_V(4,3)\right)
 \left(\tilde D_{\alpha,23,-p_2}\tilde {\bar D}_{\dot\alpha,23,-p_2}\tilde\delta_V(2,3)\right)
 \left(\tilde {\bar D}^{\dot\alpha}_{13,-p_1}\tilde\delta_V(1,3)\right),
\end{equation}
where we have used the notation (\ref{a4142}). We can readily
evaluate the covariant derivatives of the delta functions, which give
\begin{equation}
 A(1,2,3,4)={1\over256}e^{E_3^{(4)}}\theta^\alpha_{43}\bar\theta^2_{43}
 \left(\theta_{23,\alpha}\bar\theta_{23,\dot\alpha}+{1\over2}(\sigma^\mu\bar\theta_{23})_\alpha 
 p_{2,\mu}\bar\theta_{23,\dot\alpha}\theta_{23}^2\right)\bar\theta_{13}^{\dot\alpha}\theta_{13}^2.
\end{equation}
With the help of the Fierz identity (\ref{spi97}), this expression can be
rewritten as
\begin{equation}
 A(1,2,3,4)={1\over256}e^{E_3^{(4)}-p_{2,\mu}(\theta_{23}\sigma^\mu\bar\theta_{23})}
 (\theta_{43}\theta_{23})\bar\theta^2_{43}(\bar\theta_{23}\bar\theta_{13})\theta_{13}^2.\qquad
\end{equation} 
By applying this procedure to all the terms in
(\ref{a4171}), we arrive at
\begin{align}
\Gamma_{\phi^4}& (1,2,3,4)={1\over64g^2}(2\pi)^4\delta^4(p_1+p_2+p_3+p_4)
\nonumber
\\
&\times\bigg[\sin(p_1\wedge p_2)\sin(p_3\wedge p_4)\Big(F(1,2,3,4)-
 F(2,1,3,4)-F(1,2,4,3) \nonumber
\\
&\hspace*{2em} +F(2,1,4,3)+F(3,4,1,2)-F(4,3,1,2)-F(3,4,2,1)
 +F(4,3,2,1)\Big) \nonumber
\\
&~~+\sin(p_1\wedge p_3)\sin(p_4\wedge p_2)\Big(F(1,3,4,2)-
 F(3,1,4,2)-F(1,3,2,4) \nonumber
\\
& \hspace*{2em} +F(3,1,2,4)+F(4,2,1,3)-F(2,4,1,3)-F(4,2,3,1)
 +F(2,4,3,1)\Big) \nonumber
\\ 
&~~ +\sin(p_1\wedge p_4)\sin(p_2\wedge p_3)\Big(F(1,4,2,3)-
 F(4,1,2,3)-F(1,4,3,2) \nonumber
\\
& \hspace*{2em} +F(4,1,3,2)+F(2,3,1,4)-F(3,2,1,4)-F(2,3,4,1)
 +F(3,2,4,1)\Big)\bigg], \label{b4204}
\end{align}
where $F$ is given by 
\begin{equation}
 F(i,j,k,l)={1\over256}e^{E_k^{(4)}-p_{j,\mu}(\theta_{jk}\sigma^\mu\bar\theta_{jk})}
 (\theta_{lk}\theta_{jk})\bar\theta^2_{lk}\theta_{ik}^2\left(-{1\over2}\bar\theta_{ij}^2
 +{2\over3}\bar\theta_{ik}^2\right). \label{a4195}
\end{equation}
\subsection{The other vertices}
The other vertices can be evaluated in a similar way as the
$\phi^4$-vertex. Since the calculations are much easier than in the
case of the $\phi^4$-vertex, we only give the results:
\begin{align}
\Gamma_{\phi^3}(1,2,3) &= -{i\over64g^2}(2\pi)^4\delta^4(p_1+p_2+p_3)
 \textrm{sin}(p_2\wedge p_3)\nonumber
\\
& ~\times e^{p_{2,\mu}(\theta_1\sigma^\mu\bar\theta_2-
 \theta_2\sigma^\mu\bar\theta_1) 
 +p_{3,\mu}(\theta_1\sigma^\mu\bar
 \theta_3 -\theta_3\sigma^\mu\bar
 \theta_1)}\nonumber\\
& ~\times \Big({1\over8}(\theta_{21}^2-\theta_{31}^2)
(\bar\theta_{21}\bar\theta_{31})
-{1\over 8}(\theta_{21}\theta_{31})(\bar\theta^2_{21}-\bar\theta^2_{31}) 
-{1\over16}\theta_{21}^2\bar\theta_{31}^2
\nonumber
\\ 
&\qquad
+{1\over 16}\theta_{31}^2\bar\theta_{21}^2 
 -{1\over4}(p_{2,\rho}\theta_{21}\sigma^\rho
 \bar\theta_{21}-p_{3,\rho}\theta_{31}\sigma^\rho
 \bar\theta_{31})(\theta_{21}\theta_{31})
 (\bar\theta_{21}\bar\theta_{31})\Big),\nonumber
\\ \\
\Gamma_{c_-c_+\phi^2}(1,2,3,4)&= 
-{1\over384}(2\pi)^4\delta^4(p_1+p_2+p_3+p_4)\tilde\delta_V(1,2)
 \tilde\delta_V(1,3)\tilde\delta_V(1,4) \nonumber
\\
&~ \times \Big(\sin(p_1\wedge p_4)\sin(p_2\wedge p_3)
+\sin(p_1\wedge p_3)\sin(p_2\wedge p_4)\Big),
\\
\Gamma_{\bar c_-\bar c_+\phi^2}(1,2,3,4)&= 
-\Gamma_{\bar c_-c_+\phi^2}(1,2,3,4) =
 \Gamma_{c_-\bar c_+\phi^2}(1,2,3,4) = -\Gamma_{c_-c_+\phi^2}(1,2,3,4),
\\
\Gamma_{c_-c_+\phi}(1,2,3) &= 
-{i\over128}\tilde\delta_V(1,2)\tilde\delta_V(1,3)(2\pi)^4
 \delta^4(p_1+p_2+p_3)\sin(p_2\wedge p_3),
\\
 \Gamma_{\bar c_-\bar c_+\phi}(1,2,3)&= 
\Gamma_{\bar c_-c_+\phi}(1,2,3)=\Gamma_{c_-\bar c_+\phi}(1,2,3)= 
 \Gamma_{c_-c_+\phi}(1,2,3).
\end{align}
\subsection{The integrals (\ref{q1})--(\ref{q2})}
As an example, we show how to compute $I_7$ in (\ref{p2}). Inserting
the formulae for the propagators and vertices we obtain
\begin{align}
I_7 & =
-{\hbar\over i}\int dP_{V3}dP_{V4}dP_{V5}dP_{V6}\left(-{i\over128}\right)
 \tilde\delta_V(3,1)\tilde\delta_V(4,1)(2\pi)^4\delta^4(p_3+p_4+p_1)
\nonumber
\\
&\times \sin(p_4\wedge p_1)(-8)(2\pi)^4\delta^4(p_6+p_3)
 e^{p_{3,\mu}(\theta_3\sigma^\mu\bar\theta_6-\theta_6\sigma^\mu\bar\theta_3)}
 \nonumber
\\
& \times{1-(\theta_{36}\sigma^\rho\bar\theta_{36})p_{3,\rho}
+{1\over4}p_3^2\theta_{36}^2\bar\theta_{36}^2
 \over p_3^2}\left(-{i\over128}\right)\tilde\delta_V(5,2)
 \tilde\delta_V(6,2)(2\pi)^4\delta^4(p_5+p_6+p_2)\nonumber
\\
&\times\sin(p_6\wedge p_2)(-8)(2\pi)^4
 \delta^4(p_4+p_5)\,e^{p_{5,\nu}(\theta_5\sigma^\nu\bar\theta_4-\theta_4
 \sigma^\nu\bar\theta_5)}
 {1-(\theta_{54}\sigma^\lambda\bar\theta_{54})p_{5,\lambda}
+{1\over4}p_5^2\theta_{54}^2\bar\theta_{54}^2 \over p_5^2}
\nonumber
\\
& = -{\hbar\over256i}\delta^4(p_1+p_2)\int d^4k\,\partial_3^2
 \bar\partial_3^2\ldots
 \partial_6^2\bar\partial_6^2\bigg[
\sin^2(p_1\wedge k)\tilde\delta_V(3,1)\tilde\delta_V(4,1)
\nonumber
\\
&\times\tilde\delta_V(5,2)\tilde\delta_V(6,2)
 \,e^{k_{\mu}(\theta_3\sigma^\mu\bar\theta_6-\theta_6\sigma^\mu
 \bar\theta_3)}
 e^{(p_1+k)_\nu(\theta_5\sigma^\nu\bar\theta_4
-\theta_4\sigma^\nu\bar\theta_5)}
 {1\over k^2(p_1+k)^2} \nonumber
\\
&\times\left(1-(\theta_{36}\sigma^\rho\bar\theta_{36})k_{\rho}
 +{1\over4}k^2\theta_{36}^2
 \bar\theta_{36}^2\right)
\left(1-(\theta_{54}\sigma^\lambda\bar\theta_{54})(p_1+k)_\lambda
+{1\over4}(p_1+k)^2
 \theta_{54}^2\bar\theta_{54}^2\right)\bigg],\label{a4275}
\end{align}
where we have carried out the $p_4$-, $p_5$- and $p_6$-integrations
with the help of the delta functions and have renamed the remaining loop
momentum $p_3=:k$. As usual $\partial_i^2$ is a short-hand notation for
${\partial\over\partial\theta_{i,\alpha}}{
\partial\over\partial\theta_i^\alpha}$.
In the last three lines of (\ref{a4275}) we can replace $\theta_3$ and
$\theta_4$ with $\theta_1$, and $\theta_5$ and $\theta_6$ with
$\theta_2$, because of the superspace delta functions that are
contained in the integrand. Thus, we obtain
\begin{align}
I_7 & =-{\hbar\over256i}\delta^4(p_1+p_2)
 e^{-p_{1,\mu}(\theta_1\sigma^\mu\bar\theta_2
 -\theta_2\sigma^\mu\bar\theta_1)}
\int d^4k\,\sin^2(p_1\wedge k) \nonumber
\\
& \times\left[\partial_3^2 \bar\partial_3^2\ldots
 \partial_6^2\bar\partial_6^2\tilde\delta_V(3,1)\tilde\delta_V(4,1)
 \tilde\delta_V(5,2)\tilde\delta_V(6,2)\right]{1\over k^2(p_1+k)^2}
\nonumber
\\
&\times\left(1-(\theta_{12}\sigma^\rho\bar\theta_{12})k_{\rho}
 +{1\over4}k^2\theta_{12}^2 \bar\theta_{12}^2\right)
\left(1-(\theta_{12}\sigma^\lambda\bar\theta_{12})
(p_1+k)_\lambda+{1\over4}(p_1+k)^2
 \theta_{12}^2\bar\theta_{12}^2\right). \label{a4276}
\end{align}
The term between the square brackets of this expression simply gives a
factor $1$. Applying the identity (\ref{spi98}) in the last two lines
of (\ref{a4276}) we arrive at
\begin{align}
I_7&= -{\hbar\over256i}\delta^4(p_1+p_2) 
 e^{-p_{1,\mu}(\theta_1\sigma^\mu\bar\theta_2
 -\theta_2\sigma^\mu\bar\theta_1)}
\int d^4k\,\sin^2(p_1\wedge k){1\over k^2(p_1+k)^2}\nonumber
\\
&\times\left(1-(\theta_{12}\sigma^\rho\bar\theta_{12})(p_1+2k)_\rho
+ \theta_{12}^2\bar\theta_{12}^2\left({1\over4}p_1^2+p_1k+k^2
\right)\right). \label{a4277}
\end{align}
\section {Notations and Conventions}
We frequently use the definition
\begin{equation}
 {1\over2}k_i p_j \Theta^{ij}=:k\wedge p.
\end{equation}
The Fourier transform of a field is
 \begin{equation}
 \phi(x) = \int {d^4 p \over (2\pi)^4} \tilde{\phi}(p)e^{-ip x} .
 \end{equation}
The covariant derivatives are defined as 
 \begin{align}
D_\alpha &= \partial_\alpha - i(\sigma^\mu \bar\theta)_\alpha
\partial_\mu , 
&
\bar D_{\dot \alpha } &= -\bar \partial_{\dot \alpha} +
 i(\theta \sigma^\mu)_{\dot \alpha}\partial_\mu .
\end{align}
In momentum space the covariant derivatives read
\begin{align}
\tilde D_{\alpha,p} &= \partial_\alpha-p_\mu\sigma^\mu_{\alpha \dot \alpha} 
 \bar\theta^{\dot \alpha} , 
&
\tilde {\bar D}_{\dot \alpha,p} &= -\bar \partial_{\dot \alpha}+p_\mu
 \sigma^\mu_{\alpha \dot \alpha} \theta^\alpha .
\end{align}
We use the following definitions concerning Grassmann-valued objects,
\begin{align}
 \chi\eta &:=\chi^\alpha\eta_\alpha =-\eta_\alpha\chi^\alpha = \eta^\alpha
 \chi_\alpha =\eta\chi,
&
\bar\chi\bar\eta &:=\bar\chi_{\dot\alpha}\bar\eta^{\dot\alpha}=
 \bar\eta\bar\chi.
\end{align}
We have the integration measures,
\begin{align}
 \int dV &:= \int d^4x D^2 \bar D^2 , &
 \int dS &:= \int d^4x D^2 , &
 \int d\bar S &:= \int d^4x \bar D^2 .
\end{align}
The integration over $x$ cancels the total divergence parts of the covariant
derivatives. Therefore, in momentum space we have to define
\begin{align}
 \int dP_V &:=  \int {d^4p\over (2\pi)^4} \tilde D^2_{p\mapsto 0}
 \tilde {\bar D}^2_{p\mapsto 0} , 
\nonumber
\\ 
\int dP_S &:=  \int {d^4p \over (2\pi)^4} \tilde D^2_{p\mapsto 0} , 
&
\int dP_{\bar S} &:=\int{d^4p \over (2\pi)^4}  
\tilde{\bar D}^2_{p\mapsto 0} .
\end{align}
The delta functions and their Fourier transforms are given by
\begin{align}
\delta_V(1,2) &= {1 \over 16} \theta^2_{12}\bar\theta^2_{12}
\delta^4(x_1-x_2),
&
\tilde \delta_V(1,2) &= {1 \over 16} \theta^2_{12}\bar\theta^2_{12}, 
\nonumber
\\
\delta_S (1,2) &= -{1 \over 4} \theta^2_{12}\delta^4(x_1-x_2),
&
\tilde \delta_S(1,2) &= -{1 \over 4} \theta^2_{12}, 
\nonumber
\\
\delta_{\bar S}(1,2) &= -{1 \over 4} \bar \theta^2_{12}\delta^4(x_1-x_2),
&
\tilde \delta_{\bar S}(1,2) &= -{1 \over 4} \bar \theta^2_{12}.
\end{align}
Here $\theta_{ij}^\alpha := \theta^\alpha_i-\theta^\alpha_j$.
We use functional derivation in superspace:
 \begin{equation}
 {\delta_V\phi_i\over \delta_V \phi_j}= \delta^4(x_i-x_j)
 \delta_V(i,j). 
 \end{equation}
In momentum space we get an extra factor $(2\pi)^4$:
 \begin{equation}
 {\delta_V\tilde\phi_{p_i}\over\delta_V\tilde\phi_{p_j}}
 =(2\pi)^4\delta^4(p_i-p_j)\tilde \delta_V(i,j). 
 \end{equation}
Finally, we use the definition
\begin{equation}
 E_i^{(n)}:=\sum_{j=1,\ldots\!\,,n,\ j\neq i}p_{j,\mu}\left(\theta_i\sigma^\mu\bar\theta_j
 -\theta_j\sigma^\mu\bar\theta_i\right) \label{a4142}.
\end{equation}
\section {Useful Formulae}
These helpful formulae are used throughout our calculations:
\begin{align}
\theta_{ji}\theta_{ki} 
&
={1\over 2}(\theta^2_{ji}+\theta^2_{ki}-\theta^2_{jk}),
&
 (\theta\sigma^\mu\bar\theta)(\theta\sigma^\nu\bar\theta)
& ={1\over2}\eta^{\mu\nu}\theta^2
 \bar\theta^2 \label{spi98},
\\
\theta_\alpha(\theta\psi) & =-{1\over2}\theta^2\psi_\alpha ,
&
\bar\theta_{\dot\alpha}(\bar\theta\bar\psi) &=
-{1\over2}\bar\theta^2\bar\psi_{\dot\alpha}\label{spi97}.
\end{align}
\end{fmffile}
\end{document}